\documentclass[preprint,10pt]{elsarticle}

\usepackage{amssymb}
 \usepackage{amsthm}
\usepackage{graphics}
\usepackage{caption}
\usepackage{subcaption}

\begin{document}

\begin{frontmatter}



\title{Reliable SVD based Semi-blind and Invisible Watermarking Schemes}


\author[iiser]{Subhayan Roy Moulick}

\author[mathox,somerville]{Siddharth Arora}

\author[pwc]{Chirag Jain}

\author[iiser]{Prasanta K. Panigrahi}

\ead{pprasanta@iiserkol.ac.in}

\address[iiser]{Indian Institute of Science Education and Research Kolkata, India}
\address[mathox]{Mathematical Institute, University of Oxford, U.K}
\address[somerville]{Somerville College, University of Oxford, U.K.}
\address[pwc]{Price WaterHouse Coopers, India}

\begin{abstract}
A semi-blind watermarking scheme is presented based on Singular Value Decomposition (SVD), which makes essential use of the fact that, the SVD subspace preserves significant amount of information of an image and is a one way decomposition. The principal components are used, along with the corresponding singular vectors of the watermark image to watermark the target image.
For further security, the semi-blind scheme is extended to an invisible hash based watermarking scheme. The hash based scheme commits a watermark with a key such that, it is incoherent with the actual watermark, and can only be extracted using the key. Its security is analyzed in the random oracle model and shown to be unforgeable, invisible and satisfying the property of non-repudiation. 
\end{abstract}


\end{frontmatter}



\section{Introduction}
The advent of the internet has made it possible to easily store and share digital information and multimedia. While this has largely benefited all, protection of digital multimedia content has become an increasingly important issue for content owners and service providers. A common and well-proposed solution to this problem is the Digital Watermark. It is an important tool for copyright protection that embeds data into multimedia content, which can be later detected or extracted. 
Because of its key role in security and establishing ownership in multimedia files, watermarking is an area of active interest among a wide spectrum of researchers. In the last two decades considerable amount of research has been devoted to study various methods for efficiently watermarking images, which are both practically viable and theoretically sound.

Taking advantage of the optimal image decomposition property of Singular Value Decomposition for embedding a watermark in an image, several SVD based watermarking schemes have been proposed.
Singular Value Decomposition is a general linear algebraic technique, whereby a given matrix (image in this case), is diagonalized such that most of the signal energy is localized in a few singular values, \citep{GR70}. A digital image A of size M x N can be represented by its SVD as 

$$A = USV^T,$$

where $U$ and $V$ are orthogonal matrices of size $M \times M$ and $N \times N$, respectively. $S$ is a diagonal matrix of size $M \times N$, with the diagonal elements representing the singular values (SVÕs). Columns of matrix $U$, also known as left singular vectors, are the eigenvectors of $AA^T$, while columns of matrix $V$ (right singular vectors) are eigenvectors of $A^TA$. It is worth noting that, the singular vectors of an image specify the image ÔgeometryÕ, while the singular values specify the ÔluminanceÕ (energy) of the image.

For an image of size $M \times N$ (let $M>N$, without any loss of generality), the singular vectors have $O(N^2)$ elements, as compared to just $O(N)$ diagonal elements in the singular value matrix. Hence, this makes the use of singular vectors for information hiding more appropriate than using the singular values, as used by \citet{CTL05} and also by \citet{AS08}. 

It is due to the elegant properties of SVD, that it has gained much attention, and has been studied rigorously. 
\citet{LTCNTL14} used a hybrid technique, i.e., using wavelet transform and SVD followed by a recursive dither modulation algorithm, to insert signature information and textual data into the cover medical images. In addition, differential evolution was also applied to design the quantization steps optimally for controlling the strength of the watermark.
\citet{CGMJ06} divided the cover image into blocks and applied SVD to each block; the watermark was then embedded in all the non-zero singular values according to the local features of the cover image.
\citet{BD05} used SVD as a medium to embed information and give steganography scheme.
Also \citet{PSM08} proposed a SVD based watermarking and steganography schemes where watermarking scheme was designed by slightly modifying the most significant singular values for encoding vital data in the media.
\citet{L11} presented a SVD-based watermarking technique considering human visual characteristics  through block selection with Discrete Cosine Transformation (DCT) followed by SVD.
\citet{QQ04} also used DCT and SVD to construct a watermarking scheme.
\citet{ZJ11} proposed a zero-watermarking scheme combining Discrete Wavelet Transform (DWT) and SVD. The features are extracted from the cover image by applying DWT and the SVD to each non-overlapping block. The embedding of zero watermarking was realized through the exclusive or (XOR) between singular value of each block and the pixel value of the actual binary character watermark sequentially.
\citet{TJL12} gave a blind scheme based on DWT and is based on SVD and Support Vector Regression. Here, the embedding algorithm hides a watermark bit in the low-low (LL) sub-band of the cover image's principal components of the block. Additionally Particle Swarm Optimization has been utilized to optimize the scheme.
Also, \citet{RPPM15} proposed a DWT-SVD based scheme, and additionally encrypting and embedding the SVD with RSA algorithm.
\citet{PKS14} recently presented a non-blind, adaptive watermarking scheme, based on DWT and SVD, where they used principle components and perceptual tuning. 
\citet{KABA14} used a family of chaotic maps and SVD. To encrypt the watermark logo and to improve the security of watermark image Jacobian elliptic map were used, along with Quantum maps to determine the location of image's block for the watermark embedding.
Yet another commonly used method for watermarking is through Image Segmentation \citep{AAVP08}, as studied by  \citet{BKMSS02}, \citet{PSM08}, \citet{GPS11}.
For an in-depth review of Digital Watermarking and Multimedia encryption, the reader is referred to \citet{CNBH02}, \citet{FMS06}, \citet{CMBFK07}.

In this paper, we first present a SVD Based Semi-Blind watermarking scheme, whose efficacy and security depends on the fact that the SVD subspace preserves significant amount of information of an image and that SVD is a one way decomposition. 
The Semi-blind watermarking schemes (or semi-private watermarking schemes), do not require the cover image to detect the watermark. 
To introduce the watermark the principal components and the singular values of the watermark image are embedded in the cover (original) image, and the detector only needs a complimentary set for watermark extraction from the embedded watermark image.

Subsequently an Invisible Hash Based Watermarking Scheme is derived from the Semi-Blind Scheme. 
The invisible watermarking schemes, unlike visible ones, commit to a key, with embedding the watermark image in the cover (original) image, which remains incoherent with the real watermark image. The watermark can be extracted only when the original key is available. The security of the proposed invisible hash based watermarking scheme is analyzed in the random oracle model and shown to be unforgeable, invisible and satisfying the property of non-repudiation.

The paper is organized as follows. The Semi-Blind Watermarking Scheme is presented and demonstrably shown to be efficient in Section 2. In Section 3, a construction a hash code based Invisible watermarking scheme is given, and is proved to be secure in the random oracle model. Finally the practicality of the proposed schemes for multichannel (color) images are examined in section 4, followed by conclusion.

\section{Semi-Blind Watermarking Scheme}

In consideration of the fact that the singular vectors $U$ and $V$ have majority of image information, and SVD is a one way decomposition, we propose a scheme whereby the principal components of watermark image are embedded into the singular values of original image. Using this scheme, only a set of singular vectors are required to be known at the detector. The security of this emerged from the fact that SVD is a one way decomposition. 

Using SVD, a matrix, $A$, representing a mono-channel image can be represented as 

\begin{equation}\label{10}
A = USV^T,
\end{equation}

\begin{equation}\label{11}
W => U_w S_w V_w^T  => A_{wa}V_w^T,
\end{equation}

where A is the original (mono-channel) image and W is the watermark to be embedded in A. $A_{wa} = U_w$ are also known as principal components. Embedding the principal components, $A_{wa}$, with a corresponding diagonal singular value matrix S of the original image, we get $S_1$ and the corresponding watermarked image $A_w$.

\begin{equation}\label{12}
S_1 = S+ \alpha A_{wa},
\end{equation}

\begin{equation}\label{13}
U S_1 V^T => A_w,
\end{equation}

Let $A^*_w$ denote the possibly distorted watermarked image at the detector. To recover back the watermark image $W^*$ from $A^*_w$, following are the steps involved,

\begin{equation}\label{14}
(A^*_w - A) => A_1,
\end{equation}

\begin{equation}\label{15}
(U^{-1} A_1 (V^T)^{-1})/ \alpha => A^*_{wa},
\end{equation}

\begin{equation}\label{16}
A^*_{wa}V^T_w => W^*,
\end{equation}

We now try to search for a reference image, e.g., ÔPlaneÕ, in the image watermarked with baboon using our scheme. We first find the SVD of the original plane image P as,

$$P => U_p S_p V_p^T ,$$

To find a possibly distorted plane image in the distorted watermarked image $A^*_{wa}$, the singular vectors of reference image are used as follows,

\begin{equation}\label{17}
P^* => A^*_{wa} V_p^T,
\end{equation}

Since, one of the singular vectors of watermark is embedded in the original image; watermark extraction without knowing the original principal components is not possible. This clearly shows that, no reference image can be extracted from any arbitrary image using the proposed scheme [Fig.1d]. 

\begin{figure}[h!]
	\centering
	\begin{subfigure}[b] {0.45 \textwidth}
		\includegraphics[width=\textwidth]{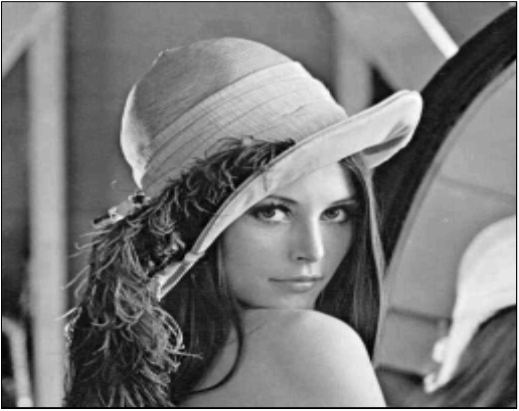}
		
		\label{fig:lena}
	\end{subfigure}		
	\begin{subfigure}[b] {0.45 \textwidth}
		\includegraphics[width=\textwidth]{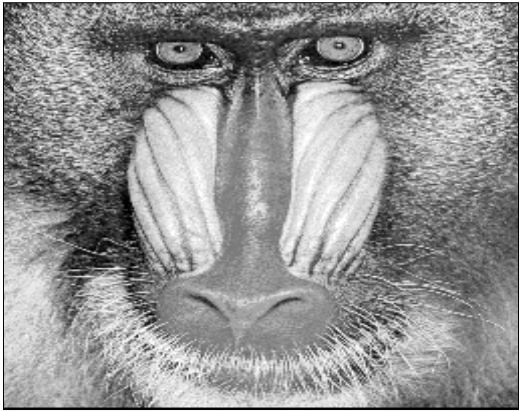}
		
		\label{fig:baboon}
	\end{subfigure}

	\begin{subfigure}[b] {0.45 \textwidth}
		\includegraphics[width=\textwidth]{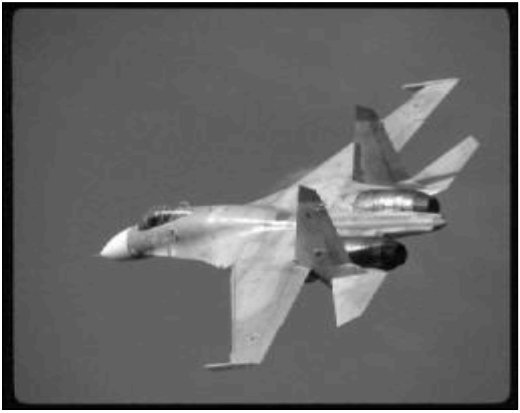}
		
		\label{fig:plane}
	\end{subfigure}
	\begin{subfigure}[b] {0.45 \textwidth}
		\includegraphics[width=\textwidth]{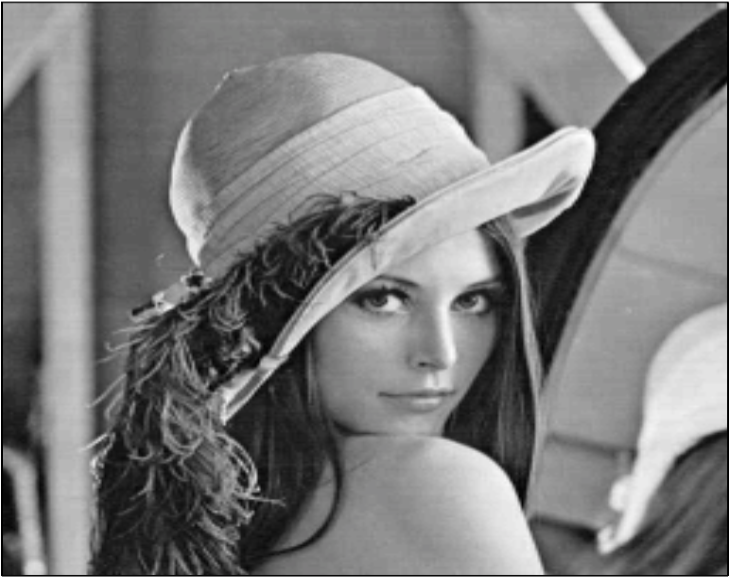}
		
		\label{fig:embed}
	\end{subfigure}

	\begin{subfigure}[b] {0.45 \textwidth}
		\includegraphics[width=\textwidth]{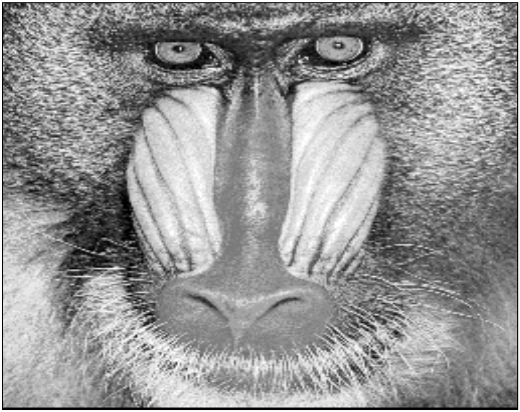}
		
		\label{fig:extracted}
	\end{subfigure}
	\begin{subfigure}[b] {0.45 \textwidth}
		\includegraphics[width=\textwidth]{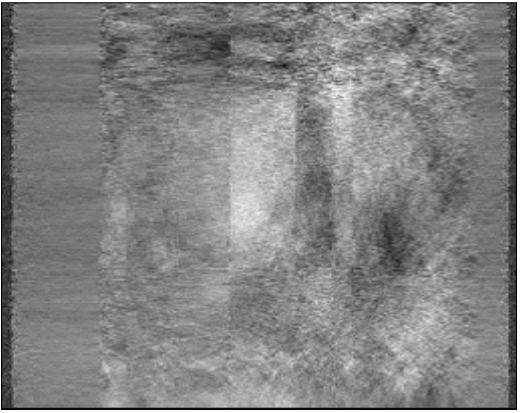}
		
		\label{fig:plane_extract}
	\end{subfigure}

	\caption{(a) Original Lena image (b) Original Baboon image (c) Original Plane image (d) Watermarked image, baboon embedded in Lena image (e) Watermark extracted from (d) using the proposed scheme (f) Distorted reference image Plane extracted from (d)}

\end{figure}

\section{A Hash Code based Invisible Watermarking Scheme}

Following the semi-blind watermarking scheme described in the preceding section, we give an invisible watermarking  scheme derived from the same. An invisible watermarking scheme essentially contains a "scrambled" watermark encoded in the original image that can be made visible only with a key. Ideally, an invisible watermarking scheme should have the following properties:
\begin{enumerate}
	\item[] \emph{Un-forgeability}: No adversary should be able to forge a watermark with an key, $id'$, that the owner (of the original image) has not watermarked, such that,
		$$ Pr[ \mathcal{A} \big( A_w, id \big) = (A_w', id')] = negligible $$
	
	\item[] \emph{Non-repudiation}: Once a signer (owner of the original image) has signed or watermarked an image with some key  $id$, the signer cannot repudiate the associate $id$, such that,
		$$ Pr[ \mathcal{A} \big(A_w, id) = (A_w, id') ] = negligible $$
	
	\item[] \emph{Invisibility}: The watermarked image, without the knowledge of the key, $id$ it was watermarked with, should be indistinguishable from an image containing a watermark of white noise from an uniform distribution, such that,
		$$ Pr [ \mathcal{D} \big(A_w \big) =1 ] - Pr [ \mathcal{D} \big(A_{U} \big) =1] = negligible$$
			
	where, $\mathcal{A}$ and $\mathcal{D}$ are polynomial time adversaries and distinguishers respectively. $A_w$ is an image watermarked with key, $id$, $A_U$ is an image containing a watermark of white noise from an uniform distribution. $id' \neq id$ and $A_w' \neq A_w$. 

\end{enumerate}

 The key idea we used here, is to commit a $M \times N$ dimensional watermark image $W$ to a unique $id$ (e.g. derived from the customer's name).  The $id$, for security reasons, must have some random nonce in it.

To do so, we use a collision resistant cryptographic hash function with polynomial span, $H$, for generating an integer string $h \in \{0, \ldots, 255\}^{M*N} \leftarrow H(id) $, and then convert $h$ to a $M \times N$ matrix $h_{id}$ using a row-major format. Informally, our hash function (of polynomial span) is defined as $H: \{0,1\}^* \rightarrow \{0,\ldots,255\}^{M*N}$, such that the following properties hold:
\begin{enumerate}
	\item It is computationally hard to find a pre-image, i.e., given $y=H(x)$, to compute $x$.
	\item It is computationally hard to find a collision, i.e., given $H(x)$, to compute $x'$ such that $H(x')=H(x)$
	\item It is indistinguishable from a random number, i.e., given $H(x)$ and $y \stackrel{R}{\leftarrow} \{0,\dots, 255\}^{|H(x)|}$, hard to distinguish w.p. $>1/2$
	 
\end{enumerate}

Finally a element-wise XOR of the matrix $A_{wa}$, as in eq. $11$ and $h_{id}$ to obtain the $A^{'}_{wa}$ is performed.

Formally, given original image $A$, a watermark $W$, and a identity $id$, we introduce a watermark to the image by first decomposing 

$$A = U S V^T, $$ 
$$W =  U_w S_w V_w^T = A_{wa} V_w^T,$$

Subsequently, we encode the watermark on to the image as 

\begin{equation}
S_1 = S + \alpha \big(A_{wa} \oplus_{e} h_{id} \big)
\end{equation}

Here $\alpha$ is the scaling factor, and $\oplus_{e}$ is the element-wise XOR, defined as $A \oplus_{e} B = A_{i,j} \oplus B_{i,j}; \forall i,j$. 

\begin{equation}
US_1 V^T = A_w,
\end{equation}

After executing the above computations, we get $A_w$ which contains a unique watermark embedded in $A$.

To extract and verify the invisible watermark, from the image $A_w$, we essentially make use of the unique id (used earlier to introduce the watermark), and to extract the watermark $W$ from $A_w$ as,

\begin{equation}
A_w - A = A_1,
\end{equation}

\begin{equation}
\big(U^{-1} A_1 {V^T}^{-1} \big)/\alpha = \big(A_{wa} \oplus_{e} h_{id} \big),
\end{equation}

\begin{equation}
\big(A_{wa} \oplus_{e} h_{id} \big) \oplus_{e} h_{id} = A_{wa},
\end{equation}

\begin{equation}
A_{wa}V_w^T = W,
\end{equation}

\subsection{Security Analysis}

\textbf{THEOREM:} If $H$ is a collision resistant secure hash function and the Semi-Blind  watermarking scheme (from section 2) is secure, then the Hash Code Watermarking Scheme is secure.

\vspace{0.2in}
\textbf{LEMMA 1:} If $H$ is a collision resistant secure hash function and the Semi-Blind  watermarking scheme (from section 2) is secure, then the Hash Code Watermarking Scheme is unforgeable.

\emph{Proof Sketch:} (by contradiction) Suppose $\exists adversary$, $\mathcal{A}$ which can forge a watermark to produce $A_w'$ using some $id'$ that the owner did not watermark. Then we can either use $\mathcal{A}$ as a subroutine of $\mathcal{B}$ that finds a collision such that $H(id) = H(id')$ or we can use $\mathcal{A}$ as a subroutine of $\mathcal{C}$ that can find a pre-image of the hash function.

It is straight forward to construct $\mathcal{B}$. $\mathcal{B}$ queries $q_{i \in poly(n)}$ to the hash oracle $\mathcal{H}$, that returns $H(q_i)$. $\mathcal{B}$ simply uses $q_i$ as key, to watermark different images and give them to $\mathcal{A}$. If $\mathcal{A}$ can produce an key $id \notin {q_i}$, such that $H(id) = H(q_i)$ for some $i$, then $\mathcal{B}$ simply returns the $H(id), id'$ to win the game. 

One can construct the pre-image finding game as follows: $\mathcal{C}$ queries $q_{i \in poly(n)}$ to the hash oracle $\mathcal{H}$, that returns $H(q_i)$.  $\mathcal{C}$ simply hash codes all watermarks with $id= q_i$ and gives them to $\mathcal{A}$. Finally when $\mathcal{A}$ returns a different $id' \neq \{ q_i \}_{\forall i}$ and possibly different $A_w$ forgery $\mathcal{C}$ computes 

$A_w' - A = A_1; U^{-1} A_1 {V^T}^{-1} = A_{wa} \oplus H(id')$

Following that, $\mathcal{C}$ computes $A_{wa} \oplus H(id') \oplus A_{wa} = H(id')$ and returns $\big(H(id'), id'\big)$ to the hash oracle to win the game. 

However, since we assumed our hash function is pre-image resistant, this is a contradiction. $\square$

\vspace{0.2in}
\textbf{LEMMA 2:} If $H$ is a collision resistant secure hash function and the Semi-Blind  watermarking scheme (from section 2) is secure, then the Hash Code Watermarking Scheme has the property of non-repudiation.

\emph{Proof Sketch:} (by contradiction) Suppose $\exists adversary, \mathcal{A}$ which can repudiate a watermark $A_w$ signed by some key $id$ with a different key $id'$, then we can use $\mathcal{A}$ as a subroutine of $\mathcal{B}$ to find a collision in the hash function. 

The construction is straightforward (as in Lemma 1) where $\mathcal{B}$ queries the hash oracle $\{q_i\}_{i\in poly(n)}$ to obtain $H(q_i), \forall i$. $\mathcal{B}$ watermarks all images with $id=q_i$ and gives them to $\mathcal{A}$. If $\mathcal{A}$ can successfully return any pair $\big(id, id'\big)$, such that $H(id)= H(id')$, $\mathcal{B}$ returns the $\big(id, id'\big)$ to the hash oracle to win the collision finding game.

However, since we assumed our hash function is collision resistant, this is a contradiction. $\square$

\vspace{0.2in}
\textbf{LEMMA 3:} If $H$ is a collision resistant secure hash function and the Semi-Blind  watermarking scheme (from section 2) is secure, then the Hash Code Watermarking Scheme is invisible.

\emph{Proof Sketch:} (by contradiction) Suppose $\exists Distinguisher, \mathcal{D}$ which can distinguish a watermark $A_w$ from $A_U$ ($A_U$ is an image contain a watermark of white noise from an uniform distribution) in polynomial time, then we can use $\mathcal{D}$ as a subroutine of $\mathcal{B}$ to win the distinguishability game. 

We construct the distinguishability game as follows: $\mathcal{B}$ queries $q_i \in poly(n)$ to the hash oracle $\mathcal{H}$, that returns $H(q_i)$. $\mathcal{B}$ then hash codes all watermark with $id= q_i$ and gives them to $\mathcal{D}$. Finally as challenge, the hash oracles gives $\big(H(id), U_r \stackrel{R}{\leftarrow} \{0, \ldots, 255 \}^{M*N} \big)$ to $\mathcal{B}$. $\mathcal{B}$ then creates two watermarks $A_w^{(1)}$ and $A_w^{(2)}$ with $H(id)$ and $U_r$ respectively and gives it to $\mathcal{D}$. $\mathcal{D}$ can distinguish the image containing the real watermark with probability more than$1/2$, and returns the correct watermark $A_w^{(b)}$  to $\mathcal{B}$. $\mathcal{B}$ can simply return the hashed string corresponding to $A_w^{(b)}$ to the hash oracle to win the game. 

However, since we assumed the output of our hash function is indistinguishable from a random sequence, this is a contradiction. $\square$

\vspace{0.2in}
\emph{Proof of Theorem:}
From \emph{Lemma 1}, \emph{Lemma 2} and \emph{Lemma 3}, we can assert that the hash code watermarking scheme is secure for any arbitrary image. $\blacksquare$

\section{Extension to Color Images}
In the previous sections, we described two constructions that introduces a watermark or fingerprint to a greyscale digital images, which are $M \times N$ matrices. This construction can be easily extended to support color images with $M \times N \times 3$ as well. One common used approach is marking only the luminance component of the the image, where the greyscale leveling techniques described in the previous sections is straightforward to apply.

The luminance component of a 3- dimensional matrix with R, G, B channels is computed as 

\begin{equation}
L = M + m 
\end{equation}
where, $M = max(R,G,B)$ and  $m = min(R,G,B)$

Yet another technique, put forward by \citet{KJB97} proposes embedding the watermark on the blue channel, since the human eye is sensitive to this band. Also another straightforward technique is to watermark each channel individually.  This can be naturally used to watermark either mono-channel or multichannel images on multichannel cover images.

\section{Conclusion}
We have presented two watermarking schemes based on SVD - a Semi-Blind Watermarking Scheme and an Invisible Hash Code based Watermarking Scheme. 
The security of the first scheme relies on the fact that the information about the entire watermark is not available without a prior knowledge of the original watermark and also the principal components along with their singular values are embedded in the watermark. These two ideas help avoid common pitfalls for the case of the semi-blind watermarking scheme.
A construction of new Invisible Hash Code Based Watermarking Scheme derived from the proposed Semi-Blind Watermarking Scheme has also been proposed here, whose security is proved in the random oracle model and shown to be unforgeable, invisible and satisfying the property of non-repudiation. 
Lastly, straightforward extensions of the aforementioned schemes for multichannel (color) images are discussed. 
Given SVD is a one way decomposition, and based on the constructions and security proofs described in the paper, we conclude that our methods help ensure rightful ownership of the digitally watermarked image. 

\biboptions{authoryear, sort&compress}
\bibliographystyle{model2-names} 

\bibliography{references}







\end{document}